\documentclass[reprint,amsmath,amssymb,aps,prb,superscriptaddress,showpacs,citeautoscript,floatfix]{revtex4-1}
\usepackage{graphicx}
\usepackage{dcolumn}
\usepackage{bm}
\usepackage{amsmath}
\usepackage{amssymb}
\usepackage[]{units}
\usepackage{subcaption}
\usepackage{booktabs}
\usepackage{lineno}
\begin{document}

\title{Adhesion and material transfer between contacting Al and TiN surfaces from first principles}

\author{G. Feldbauer}
\email{gf@cms.tuwien.ac.at}
\affiliation{Institute of Applied Physics, Vienna University of Technology, Gu\ss hausstra\ss e 25-25a, 1040 Vienna, Austria}
\affiliation{Austrian Center of Competence for Tribology, AC2T research GmbH, Viktor-Kaplan-Stra\ss e 2, 2700 Wiener Neustadt, Austria}
\author{M. Wolloch}
\affiliation{Institute of Applied Physics, Vienna University of Technology, Gu\ss hausstra\ss e 25-25a, 1040 Vienna, Austria}
\affiliation{Austrian Center of Competence for Tribology, AC2T research GmbH, Viktor-Kaplan-Stra\ss e 2, 2700 Wiener Neustadt, Austria}
\author{P. O. Bedolla}
\affiliation{Institute of Applied Physics, Vienna University of Technology, Gu\ss hausstra\ss e 25-25a, 1040 Vienna, Austria}
\affiliation{Austrian Center of Competence for Tribology, AC2T research GmbH, Viktor-Kaplan-Stra\ss e 2, 2700 Wiener Neustadt, Austria}
\author{P. Mohn}
\affiliation{Institute of Applied Physics, Vienna University of Technology, Gu\ss hausstra\ss e 25-25a, 1040 Vienna, Austria}
\author{J. Redinger}
\affiliation{Institute of Applied Physics, Vienna University of Technology, Gu\ss hausstra\ss e 25-25a, 1040 Vienna, Austria}
\author{A. Vernes}
\affiliation{Institute of Applied Physics, Vienna University of Technology, Gu\ss hausstra\ss e 25-25a, 1040 Vienna, Austria}
\affiliation{Austrian Center of Competence for Tribology, AC2T research GmbH, Viktor-Kaplan-Stra\ss e 2, 2700 Wiener Neustadt, Austria}

\begin{abstract}
A series of density functional theory (DFT) simulations was performed to investigate the approach, contact, and subsequent separation of two atomically flat surfaces consisting of different materials. Aluminum (Al) and titanium nitride (TiN) slabs were chosen as a model system representing a metal-ceramic interface and the interaction between soft and hard materials. The approach and separation were simulated by moving one slab in discrete steps normal to the surfaces allowing for electronic and atomic relaxations after each step. Various configurations were analyzed by considering (001), (011), and (111) surfaces as well as several lateral arrangements of these surfaces at the interface. Several tests were conducted on the computational setup, for example, by changing the system size or using different approximations for the exchange correlation functional. The performed simulations revealed the influences of these aspects on adhesion, equilibrium distance, and material transfer. These interfacial properties depend sensitively on the chosen configuration due to distinct bond situations. Material transfer, in particular, was observed if the absolute value of the adhesion energy for a given configuration is larger than the energy cost to remove surface layers. This result was found to be independent of the employed exchange correlation functional. Furthermore, it was shown that a simple comparison of the surface energies of the slabs is not sufficient to predict the occurrence of material transfer.
\end{abstract}


\maketitle

\section{Introduction}\label{sec:intro}
Contacts of surfaces at the atomic length scale are crucial in many modern applications, from experimental techniques such as nanoindentation~\cite{landman_atomistic_1990, oliver_measurement_2004, fischer-cripps_nanoindentation_2004} or atomic/friction force microscopy (AFM/FFM)~\cite{binnig_atomic_1986, meyer_scanning_2004, bennewitz_friction_2005} to nanotechnologies applied, for example, in nano-/microelectromechanical-systems (NEMS/MEMS)~\cite{komvopoulos_surface_1996, spearing_materials_2000, maboudian_surface_2004, kim_nanotribology_2007, bhushan_nanotribology_2008}. The reliability, performance, and lifetime of such systems, for example, depend sensitively on the interactions between contacting materials. Furthermore, detailed insights into such contacts are of fundamental interest for better comprehension of tribological processes, such as nanoscale wear~\cite{bhushan_nanotribolgy_1995, gnecco_abrasive_2002, bhushan_nanotribology_2005, gnecco_fundamentals_2007, gotsmann_atomistic_2008, bhaskaran_ultralow_2010, jacobs_nanoscale_2013, mishina_wear_2013}, for which there is still a lack of understanding due to its highly complex nature~\cite{kim_nano-scale_2012}. 

Metal-ceramic interfaces~\cite{howe_bonding_1993} are of fundamental and technological interest because they exhibit advantages of both types of materials such as valuable mechanical properties, high thermal stability, and degradation resistance~\cite{johansson_electronic_1995}. Hence, such interfaces are important in numerous applications such as communication devices and nanoelectronics~\cite{ruhle_preface_1992}. In this paper the interface between the metal Al and the transition-metal nitride TiN is investigated. This interface consists of a soft material and a hard material, which simplifies wear processes because the softer material is primarily affected.

Since the 1980s classical molecular dynamics (MD) simulations have commonly been applied to nanotribological problems (see, e.g., Refs.~\onlinecite{thompson_simulations_1989, landman_structural_1989, bhushan_computer_2000, mulliah_molecular_2004, kenny_molecular_2005, david_schall_molecular_2007, szlufarska_recent_2008, vernes_three-term_2012, eder_derjaguin_2013, eder_methods_2014, eder_analysis_2014}) and still constitute a standard tool in numerical atomistic simulations. Nevertheless, during the last decade density functional theory (DFT) calculations have been increasingly used in this field (see, e.g., Refs.~\onlinecite{zhong_first-principles_1990, dag_atomic_2004, ciraci_ab-initio_2007, zilibotti_ab_2009, garvey_shear_2011, zilibotti_ab_2011, cahangirov_frictional_2012, kwon_enhanced_2012, wang_theoretical_2012, wang_atomic-scale_2012, garvey_pressure_2013, wolloch_ab-initio_2014}) and should be seen as an extension to the more common computational tools in tribology. DFT allows for parameter-free calculations and an accurate description of quantum-mechanical systems and does not depend on empirical potentials. However,  due to computational challenges DFT calculations are currently limited to rather small systems of typically a few hundred atoms. Since DFT has proven to yield reliable results for this class of systems~\cite{finnis_theory_1996, lundqvist_density-functional_2001, sinnott_ceramic/metal_2003}, it is also employed in this study to analyze the electronic and atomic structure of the Al/TiN interfaces and to determine properties such as adhesion energies. Results obtained with DFT, such as potential-energy curves, can be used as an input for, e.g., large-scale classical MD simulations~\cite{ercolessi_interatomic_1994, jaramillo-botero_general_2014}. Furthermore, the combination of approaches such as DFT and MD as well as the continuously increasing available computer power and advances in software tools promise the possibility to investigate even larger and more realistic systems in the near future. 

Al/TiN and similar interfaces have already been investigated by various researchers with experimental~\cite{avinun_nucleation_1998, chun_interfacial_2001-1} and theoretical~\cite{liu_adhesion_2003, liu_first-principles_2004, liu_first-principles_2005, song_adhesion_2006, zhang_first-principles_2007, zhang_effects_2007, song_mechanism_2008, yadav_first-principles_2012, yadav_first-principles_2014} methods. Here, however, the emphasis lies on a realistic way to simulate the separation of the interfaces as well as on a comprehensive discussion of interfaces between Al and TiN low-index surfaces. To assess this problem, the effects of various configurations at the interface as well as approach and subsequent separation of Al and TiN slabs are analyzed. Various tests on, e.g., the effect of adjusted lattice parameters, the simulation cell size, and various approximations for the exchange correlation functional in DFT are carried out.

\section{Computational Details}\label{sec:comp_methods}

\subsection{Density Functional Theory Calculations}

To study the interfacial properties of Al and TiN slabs upon approach and subsequent separation, we performed first-principles calculations within the framework of DFT employing the Vienna Ab initio Simulation Package (\textsc{VASP})~\cite{kresse_ab_1993, kresse_ab_1994, kresse_efficient_1996, kresse_efficiency_1996}. \textsc{VASP} utilizes a plane-wave basis and periodic boundary conditions.  Projector augmented-wave (PAW) pseudopotentials~\cite{blochl_projector_1994,kresse_ultrasoft_1999} were used to model the potential between the ionic core and the valence electrons. Unless explicitly mentioned otherwise, the generalized gradient approximation (GGA) in the Perdew, Burke, and Ernzerhof (PBE) parametrization~\cite{perdew_generalized_1996} was applied to describe the exchange and correlation functional. Since GGAs often underestimate binding and adhesion energies~\cite{stampfl_density-functional_1999}, the local-density approximation (LDA)~\cite{perdew_self-interaction_1981}, which usually overestimates these quantities~\cite{van_de_walle_correcting_1999}, was also employed for comparison. Additionally, the van der Waals (vdW) density functional (DF) optB86b~\cite{klimes_chemical_2010,klimes_van_2011} was used, which includes a nonlocal correlation term approximating vdW interactions. vdW-DFs have been applied to a wide range of materials (e.g., see Refs.~\onlinecite{chakarova-kack_application_2006, sony_importance_2007, carrasco_wet_2011, mittendorfer_graphene_2011, antlanger_pt_3zr0001:_2012, graziano_improved_2012, bedolla_effects_2014, choi_growth_2014, bedolla_density_2014}) and have proven to be of good accuracy. Although vdW interactions should not play a major role in the investigated systems, the calculations are included for comparison and clarification. The calculation parameters were carefully chosen to obtain accurate total energies. An energy cutoff of \unit[800]{eV} was used for the plane-wave basis. Unless noted otherwise, the Brillouin zone sampling was performed using a \(\Gamma\)-centered \(15\times15\times1\) Monkhorst-Pack mesh~\cite{monkhorst_special_1976}. Both settings  allow for total energies accurate to \unit[1]{meV/atom}. While the tetrahedron method with Bl\"ochl corrections~\cite{blochl_improved_1994} was utilized for static calculations, for relaxations a smearing of \unit[0.11]{eV} using the first-order method of Methfessel and Paxton~\cite{methfessel_high-precision_1989} was selected. In order to relax the structures a damped molecular dynamics (MD) algorithm was employed, allowing for atomic movements until an energy convergence criterion of \unit[\(10^{-5}\)]{eV} was fulfilled. This damped MD scheme was chosen instead of the widely used quasi-Newton or conjugate-gradient algorithms, because these caused convergence problems as well as the tendency to remain stuck in local minima. Each converged relaxation was followed up by a static calculation to obtain more accurate total energies. For electronic self-consistency cycles a convergence criterion of \unit[\(10^{-6}\)]{eV} was used. All simulations were performed at \unit[0]{K}.

\subsection{Simulation Model}\label{subsec:sim-model}

\begin{figure}[hbt]
  \centering
  \includegraphics[width=.25\linewidth]{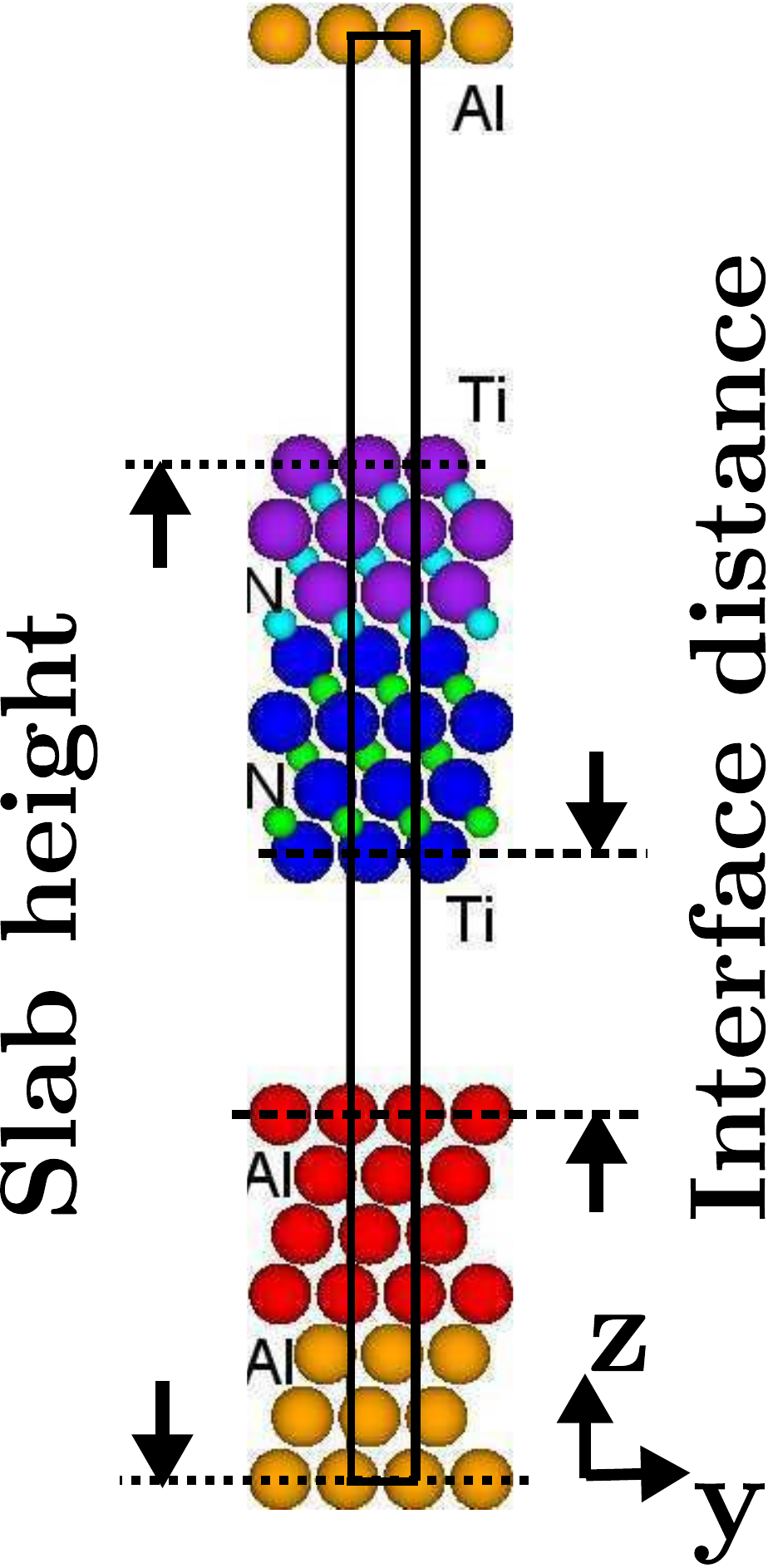}
  \caption{Side view of a (111) Al/TiN interface (TiN: Ti terminated). The simulation interface cell is indicated by the solid black lines. During relaxations the orange Al, cyan N and purple Ti atoms were kept rigid, while the red Al, green N and blue Ti ones were allowed to relax.}
  \label{fig:Al-TiN-initial}
\end{figure}

To model our systems we built simulation cells from a fcc Al slab at the bottom and a rock salt TiN slab above (see Fig.~\ref{fig:Al-TiN-initial}). Such cells were constructed for the low-index surface orientations (001), (011), and (111) of both slabs. Only configurations with slabs of the same surface orientations at the interface and without any relative rotations were considered. The two slabs were separated by a gap which is given by the vertical distance between the top Al and bottom TiN layers and will be referred to as the ``interface distance''. The vertical distance between the bottom Al and top TiN layers, which is the sum of the interface distance and the heights of the two slabs,  is called ``slab height''. In the case of (111) slabs this height is measured up to the top Ti and N layer for Ti and N termination, respectively. Unless otherwise stated 1\(\times\)1 surface cells were used, which represent an infinitely extended surface due to the periodic boundary conditions. The Al slab consisted of at least seven layers, and the TiN slab consisted of a minimum of six Ti and six N layers. These thicknesses were found to be sufficient to converge the surface energies and to mimic bulklike features in the center of the respective slab. These system dimensions are in good agreement with other published work~\cite{marlo_density-functional_2000, liu_first-principles_2004, zhang_first-principles_2007, yadav_first-principles_2014}. 

\begin{figure}[hbt]
  \centering
  \includegraphics[width=.9\linewidth]{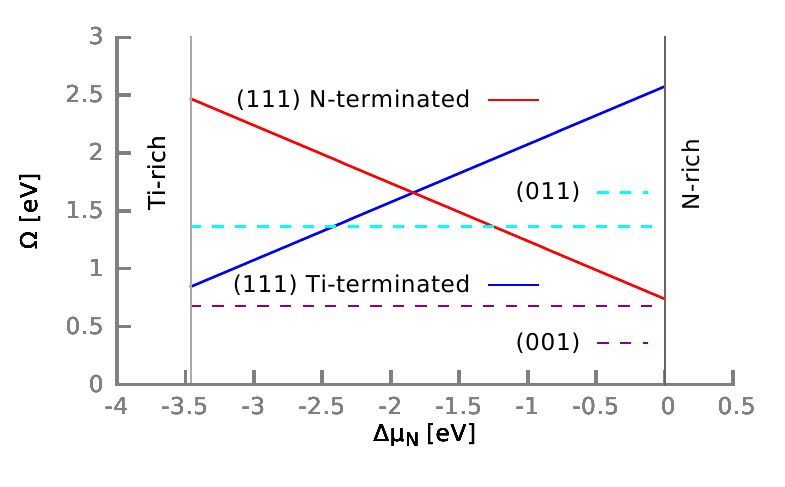}
  \caption{Surface phase diagram for TiN. The surface Gibbs free energy \(\Omega\) [see Eq.~\eqref{equ:surf-energy}] referenced to a 1\(\times\)1 surface cell of the (111) orientation is plotted vs the deviation \(\Delta\mu_{N}\) of the nitrogen chemical potential from its molecular reference [see Eq.~\eqref{equ:delta-mu}] for N- and Ti-terminated (111) TiN slabs (solid lines) as well as for (001) and (011) orientations (dashed lines).}
  \label{fig:TiN-term}
\end{figure}

The (111)~TiN slab can be terminated with either Ti or N atoms. To investigate the stability of these terminations a thermodynamic analysis~\cite{wang_hematite_1998, reuter_composition_2001} was performed to calculate the surface Gibbs free energy for the off-stoichiometric slabs~\cite{lee_stoichiometry_2011}. The surface Gibbs free energy \(\Omega\) for surface termination \(i\) without vibrational contributions is given by
\begin{equation}
  \Omega^i = \frac{1}{2} \left( E^i_{slab} - N^i_{Ti} E^{bulk}_{TiN} \right) -  \Gamma^i_{Ti,N} E_N - \Gamma^i_{Ti,N} \Delta\mu_N,
  \label{equ:surf-energy}
\end{equation}
where \(E^i_{slab}\) is the total energy of the slab with termination \(i\), \( N^i_{Ti}\)  isthe number of Ti atoms in the slab, \(E^{bulk}_{TiN}\) is the total energy of bulk TiN, and \(E_N\) is the total energy of a nitrogen atom. The two latter terms in Eq.~\eqref{equ:surf-energy} are necessary to calculate the surface energy of off-stoichiometric slabs. The number of off-stoichiometric atoms \(\Gamma^i_{Ti,N}\) is defined as
\begin{equation}
  \Gamma^i_{Ti,N} = \frac{1}{2} \left( N^i_N - N^i_{Ti} \frac{N^{bulk}_N}{N^{bulk}_{Ti}} \right),
  \label{equ:gamma}
\end{equation}
where \(N^i_j\) and \(N^{bulk}_j\) are the number of atoms of type j in the slab and in bulk, respectively. For rock-salt bulk TiN the fraction \(N^{bulk}_N/N^{bulk}_{Ti}\) in Eq.~\eqref{equ:gamma} is equal to 1. \(\Delta\mu_N\) is the deviation of the nitrogen chemical potential \(\mu_N\) from the molecular reference \(\frac{1}{2} E_{N_2}\),
\begin{equation}
  \Delta\mu_N = \mu_N - \frac{1}{2} E_{N_2}.
  \label{equ:delta-mu}
\end{equation}
In Figure~\ref{fig:TiN-term} the calculated surface Gibbs free energy is plotted for the N- and Ti-terminated TiN (111) slabs in the stability range of nitrogen in TiN obtained from the heat of formation of bulk TiN~\cite{chase_nist-janaf_1998} at \unit[0]{K}, \(\Delta H^0_{f,(TiN)} =-3.461\)~\unit[]{eV}, and the chemical potential of gas phase nitrogen, i.e., \( \Delta H^0_f \le \Delta\mu_N \le 0\). Fig.~\ref{fig:TiN-term} shows that the favorable termination of a (111) TiN slab depends on the chemical potential of nitrogen, in agreement with the Refs.~\onlinecite{liu_first-principles_2004, wang_surface_2010}. Since both cases are found in reasonable nitrogen concentration ranges, both terminations are  investigated.

Dipole corrections~\cite{neugebauer_adsorbate-substrate_1992} perpendicular to the interface (z direction) were tested for the systems but were found to be negligible. Atop the TiN slab a vacuum spacing of at least \unit[10]{\AA} was included to decouple periodically repeated cells in the z direction. The lattice parameters of the single slabs, \unit[4.04]{\AA} and \unit[4.254]{\AA} for Al and TiN, respectively, were obtained from bulk calculations. These values are in very good agreement with the experimental lattice constants of \unit[4.05]{\AA} and \unit[4.265]{\AA} for Al and TiN, respectively~\cite{wyckoff_crystal_1963}. The relative error between calculated and experimental values is below 0.5\%. For the simulation cells combining Al and TiN slabs, unless otherwise stated, an intermediate lattice parameter of \unit[4.144]{\AA} was used for the lateral xy lattice vectors to equalize the relative error of about 2.6\% for both materials. For the z direction the material-specific values were kept assuming a pseudomorphic interface. In reality such a combination of stretching and compression of thick slabs does not usually occur, but dislocations at the interface or an incommensurate contact are possible. Thus, some of the atoms on both sides of the interface would not be aligned perfectly, but rather sample slightly different local environments. For computational reasons, here these different local arrangements are assessed by considering various orientations at the interface as limiting cases.

The approach of the two slabs was simulated by moving the upper slab in discrete steps along the negative z direction and allowing for electronic and atomic relaxations after each step. Alternatively, moving the bottom slab toward the upper slab or both toward each other would not affect the results. For the atomic relaxations the top TiN (three Ti and three N) and the bottom three Al layers were kept fixed at bulklike distances, whereas the intermediate ``free'' ones were allowed to fully relax. This is depicted in Fig.~\ref{fig:Al-TiN-initial} for the Ti-terminated (111) surface orientation. For the approaching movement a step size of \unit[0.2]{\AA} was used for all configurations. Before the slabs were brought into contact, the free layers were allowed to relax in order to simulate surfaces in their equilibrium for the chosen lattice parameters. The separation of the slabs was initiated from the equilibrium, i.e., the structure with the lowest energy determined during the approach. To simulate a realistic separation of the slabs only the topmost, rigid TiN layers were moved in discrete steps in the positive z direction, again allowing for electronic and atomic relaxations after each step. The choice of the step size is crucial for the separation process. Separation velocities allowing for an adiabatic behavior of the system were assumed, meaning that the system continuously fully adjusts during the separation at each step. However, this assumption should also be valid for velocities up to several hundred meters per second as long as these are still considerably lower than the material-specific speed of sound, which is above \unit[6000]{m/s} for Al and TiN~\cite{kundu_ultrasonic_2012}. It is evident that the step size has to be small enough to mimic the adiabatic relaxation, but on the other hand, a smaller step size leads to increased computational costs. For the investigated systems a step size of \unit[0.1]{\AA} was found to be a practical trade-off because calculations showed this value to be necessary to converge the final results of the simulated separation processes. Smaller step sizes down to \unit[0.01]{\AA} were also considered for approach and separation but did not yield qualitatively different results. Clearly, quantities such as the slab height corresponding to the initial material transfer can be determined more accurately.

\begin{figure}[hbt]
  \centering
  \includegraphics[width=1.\linewidth]{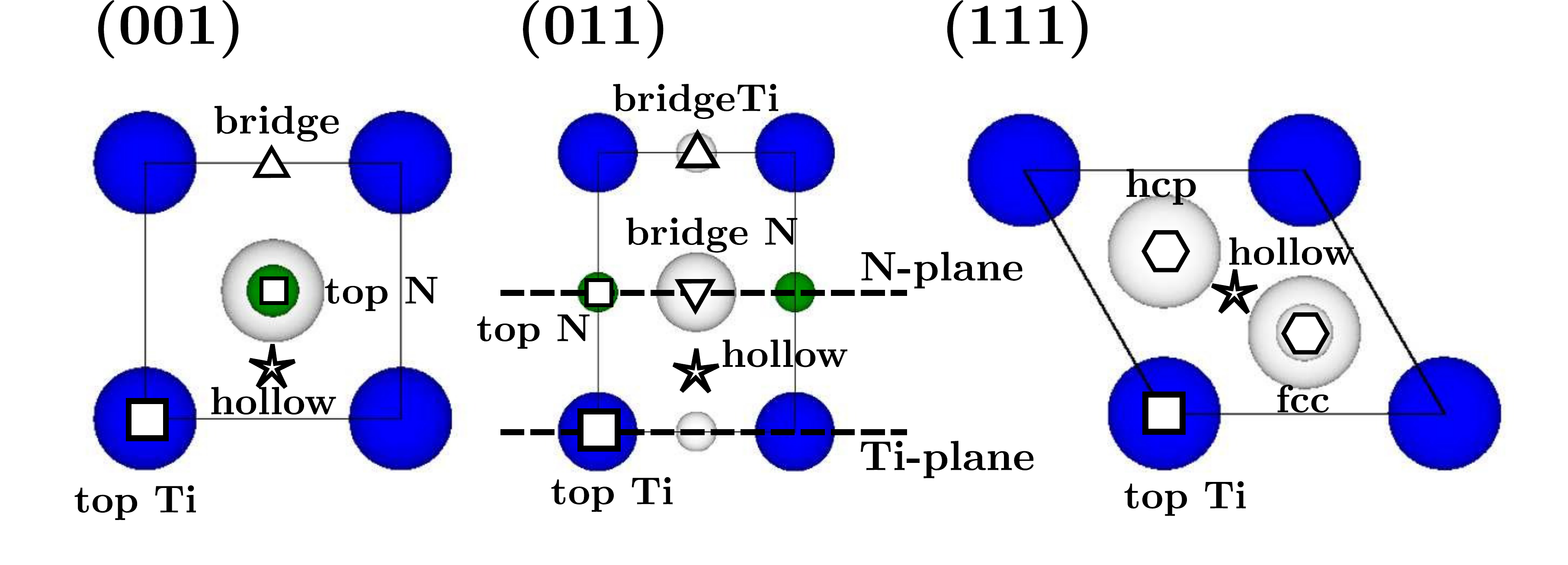}
  \caption{Top view of (001), (011), and Ti-terminated (111) TiN surfaces. For each orientation the 1\(\times\)1 surface cell is presented. Filled circles indicate atoms in the top surface layer (Ti and N are given by large blue and small green circles, respectively), while empty circles label atoms below the top surface layer. To obtain a N-terminated (111) TiN surface the Ti and N atoms of the Ti-terminated surface have to be exchanged. High-symmetry points are highlighted. For the (011) TiN surface the ``Ti plane'' and ``N plane'' are marked by dashed lines.}
  \label{fig:bond-sites}
\end{figure}

In order to study the effects of different alignments of the slabs at the interface, the upper slab was also laterally placed on various sites with respect to the surface of the lower slab. The definitions of the configurations are depicted in Fig.~\ref{fig:bond-sites} by marking the high-symmetry points on the low index TiN surfaces where the next Al atom can be placed. In this context the interaction energy \(E_I(z)\) is an important quantity, which is defined as the difference of the total energy of the interacting slabs \(E_{(Al/TiN)}(z)\) at slab height \(z\) and the reference energies  of the independent slabs, \(E_{(Al)}\) and \(E_{(TiN)}\),
\begin{equation}
  E_I(z) = E_{(Al/TiN)}(z) - E_{(Al)} - E_{(TiN)}.
  \label{equ:interaction-energy}
\end{equation}

\section{Results and Discussion}\label{sec:results}

\subsection{Removal of Layers from an Al Slab}\label{subsec:remove-al}

As a first step the energy cost for removing layers from an Al slab was examined for all three low-index surface orientations. The removal of the layers was simulated by placing the layers at a large distance from the slab, which does not allow for interactions between the slab and layers. The TiN slab is not investigated here because the Al slab is assumed to be mainly affected by deformations or material transfer within an Al/TiN interface because TiN forms a much more rigid lattice. The energetical results for the removal of the top Al layer are given in Table~\ref{tab:removal}. These removal energies are calculated for simulation cells using the bulk lateral lattice parameters as well as the modified ones used for the Al/TiN simulation cell. For the modified lattice parameters the removal energies are typically overestimated by about 5\%--10\%, meaning that it is actually easier to remove layers from the equilibrium structure. The removal energy for the modified Al slab is increased because the lateral stretching causes a vertical compression of the slab if relaxations are allowed. This compression occurs to minimize the volume change and locally strengthens the bonding of the surface layers. This effect is strongest for the top surface layer, which moves about \unit[0.24]{\AA} towards the rigid part and becomes weaker for the subsurface layers; for example, the fourth layer is only shifted by about \unit[0.08]{\AA}. 

\begin{figure}[hbt]
  \centering
  \includegraphics[width=0.9\linewidth]{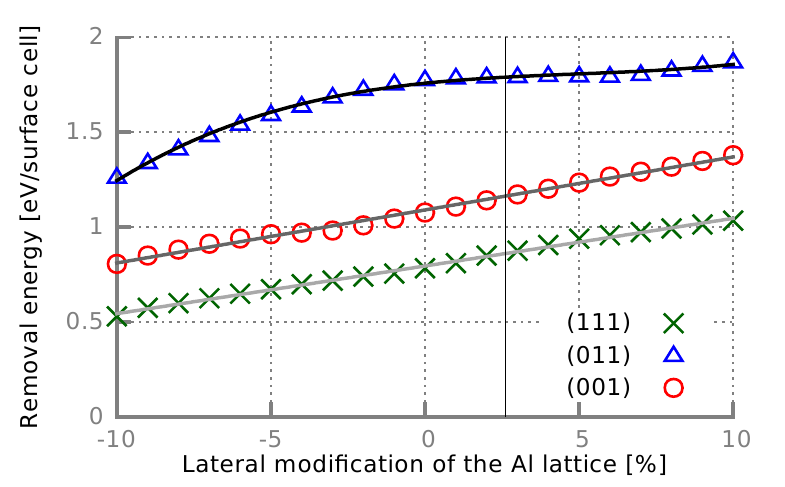}
  \caption{PBE energy costs to remove the top Al layer for the (001), (011), and (111) surface orientations. The lateral effects of stretching and compression of the 1\(\times\)1 surface cell on the removal energies are shown. The Al bulk lattice constant is used as a reference value at 0\%. The vertical line indicates the intermediate Al/TiN interface lattice parameter, while the other solid lines are given to guide the eye.}
  \label{fig:Al-xy-removal}
\end{figure}

The influence of compressive and tensile stress on the removal energies of the top Al layer is illustrated in Fig.~\ref{fig:Al-xy-removal} for the three low-index surface orientations. The data points for the (001) and (111) surfaces follow a similar trend, whereas the behavior of the (011) surface clearly deviates. This difference occurs probably due to the openness of the (011) surface and the significant impact of relaxations. The influence of stress, found for all surfaces, supports the notion of stress-assisted wear~\cite{gotsmann_atomistic_2008,jacobs_nanoscale_2013}, which states the possibility of a reduction of the activation barriers for the detachment of atoms from a structure due to stress. Furthermore, different approximations for the exchange correlation functional were tested. As expected it was found that LDA and the vdW-DF optB86b yield larger removal energies by about 15\%--20\%, where LDA typically gives values larger by a few percent than the vdW-DF (see Table~\ref{tab:removal}).
\begin{table}[hbt]
  \caption{\label{tab:removal}  Energy costs to remove the top layer from an Al slab for the (001), (011), and (111) surface orientations using PBE, LDA, and vdW-DF optB86b. The removal energies are given in \unit{eV} per 1\(\times\)1 surface cell. \(a_{Al}\) is the Al bulk lattice parameter, whereas \(a_{Al/TiN}\) corresponds to the modified Al/TiN interface.}
  \begin{ruledtabular}
    \begin{tabular}{lccc}
      & (001) & (011) & (111) \\
      PBE (\(a_{Al}\)) & 1.08  & 1.78  & 0.78 \\
      PBE (\(a_{Al/TiN}\)) & 1.16  & 1.79  & 0.87  \\
      LDA (\(a_{Al/TiN}\)) & 1.33  & 2.02  & 1.02  \\
      vdW (\(a_{Al/TiN}\)) & 1.27  & 1.89  & 1.00 
    \end{tabular}
  \end{ruledtabular}
\end{table}

\subsection{Lateral Alignments at the Al/TiN Interface}\label{subsec:alignments}

\begin{figure*}[hbt]
  \centering
  \begin{subfigure}[b]{0.45\linewidth}
    \includegraphics[width=\linewidth]{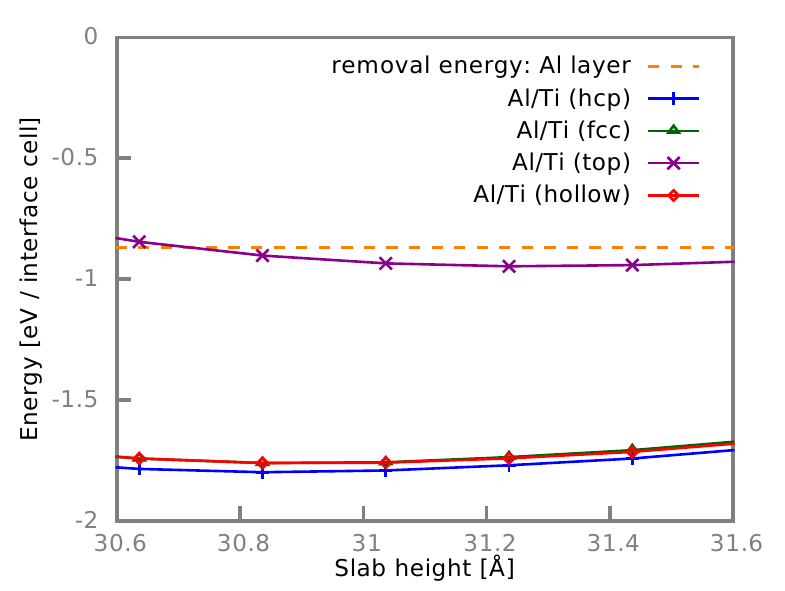}
    \caption{(111) Ti-terminated}
    \label{fig:111-Titerm-bond-site}
  \end{subfigure}
  \begin{subfigure}[b]{0.45\linewidth}
    \includegraphics[width=\linewidth]{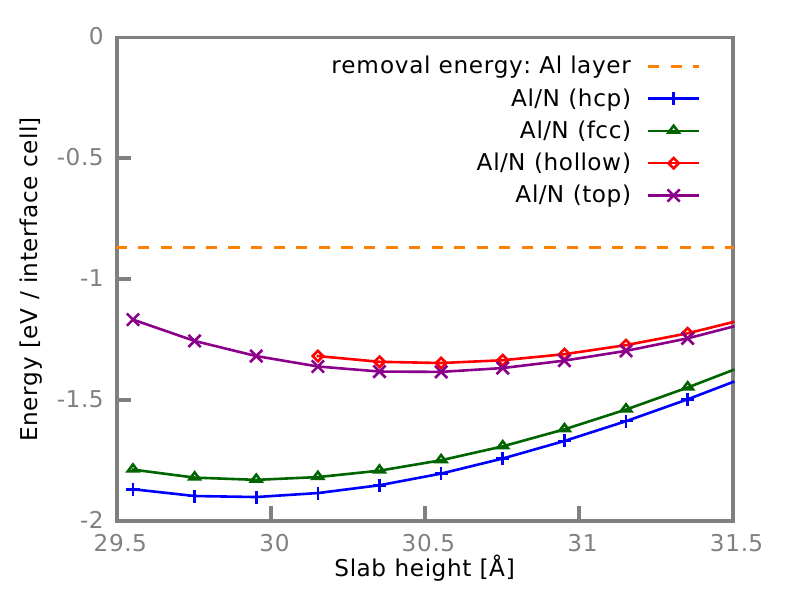}
    \caption{(111) N-terminated}
    \label{fig:111-Nterm-bond-site}
  \end{subfigure}
  \begin{subfigure}[b]{0.45\linewidth}
    \includegraphics[width=\linewidth]{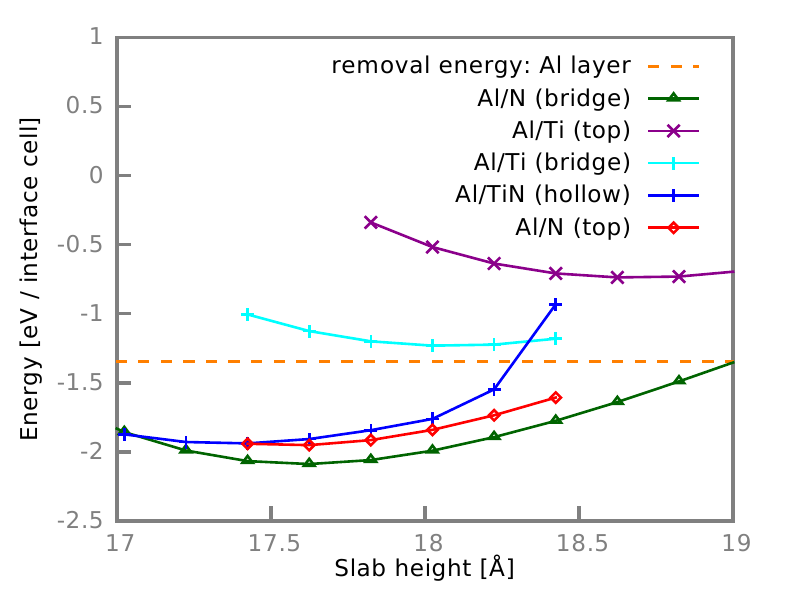}
    \caption{(011)}
    \label{fig:110-bond-site}
  \end{subfigure}
  \begin{subfigure}[b]{0.45\linewidth}
    \includegraphics[width=\linewidth]{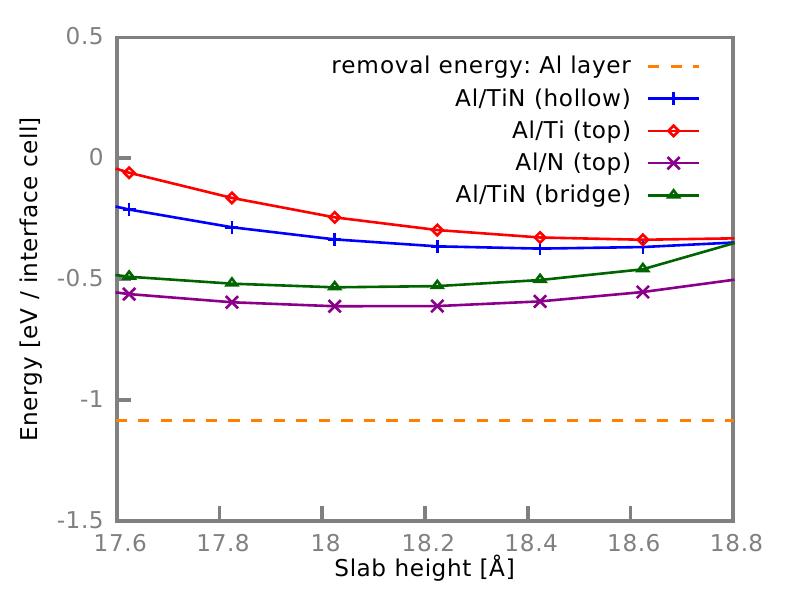}
    \caption{(001)}
    \label{fig:100-bond-site}
  \end{subfigure}
  \caption{Calculated PBE interaction energies of the relaxed Al/TiN interface for the (111) Ti-terminated, (111) N-terminated, (011), and (001) surface orientations. Various lateral alignments of the two slabs are considered (see Fig.~\ref{fig:bond-sites}). The horizontal, dashed orange lines give the energy costs to remove at least one layer from an Al slab of the corresponding surface orientation.}
  \label{fig:bond-sites-pec}
\end{figure*}

Effects of various lateral alignments of the slabs at the interface (see Fig.~\ref{fig:bond-sites}) were investigated for the different surface orientations. These studies revealed the strong dependence of equilibrium properties such as adhesion energies and the equilibrium distances on the chosen configuration. The calculated interaction energies~[Eq.~\eqref{equ:interaction-energy}] of relaxed interfaces are shown in Figs.~\ref{fig:bond-sites-pec}(a)--(d) for slab heights around the energy minima, which are equivalent to the adhesion energies for each alignment. In general, the top Al atoms prefer the proximity of N atoms over Ti. The bonding situation will be discussed in more detail in the following paragraphs. From an energetical point of view material transfer between the slabs should be possible only if the energy cost to remove layers is compensated for. Thus, the energy gain due to adhesion has to be larger than the energy cost to remove one or more layers. This argument is sketched in Figs.~\ref{fig:bond-sites-pec}(a)--(d) by including a horizontal line at the negative value of the Al removal energy for each surface orientation. It has been observed experimentally that metal-ceramic interfaces with weak and strong interfacial adhesion break upon stress at the interface and in bulk areas, respectively~\cite{howe_bonding_1993-1, ernst_metal-oxide_1995}. We find that the four surfaces investigated exhibit essentially different behavior. The adhesion energies and the equilibrium distances, i.e., the interface distances at the minimum of each energy curve, depend strongly on the surface orientation as well as on the alignment at the interface. In the case of the (111) surfaces all configurations should lead to the removal of at least one Al layer. For the (011) surfaces this is the case only for three alignments (see Fig.~\ref{fig:bond-sites}), Al/N~(top), Al/TiN~(hollow), and Al/N~(bridge). In contrast, for the (001) surfaces no material transfer should occur since for all cases studied the energy to remove one Al layer is larger than the adhesion energy. 

As mentioned above, in reality, surfaces with a bulk lattice mismatch are usually not perfectly aligned at an interface. Consequently, not all atoms are placed on the same contact site; therefore, the interfacial properties such as the adhesion energy are an average of the actually occupied sites. The configurations presented here, however, constitute limiting cases of perfectly aligned systems, such that the properties of real interfaces should be found within these boundaries. 

\begin{figure*}[hbt]
  \centering
  \begin{subfigure}[b]{0.45\linewidth}
    \includegraphics[width=\linewidth]{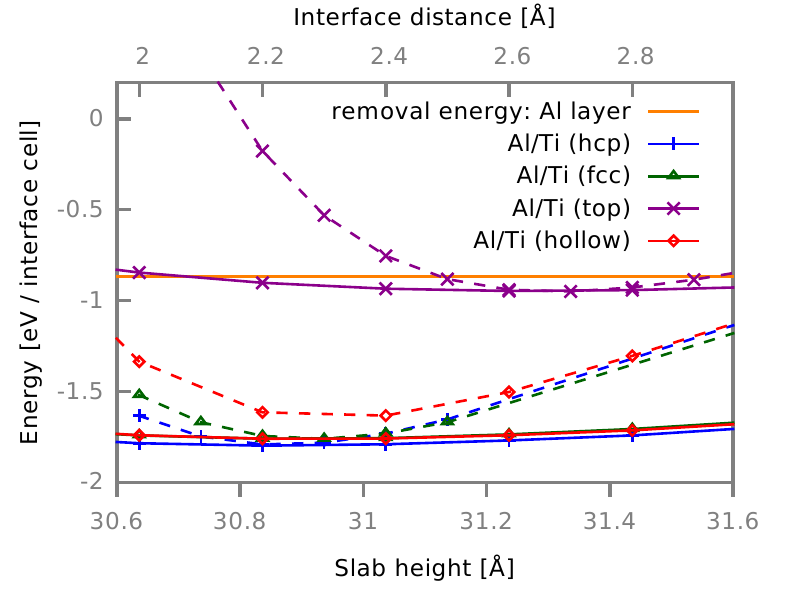}
    \caption{(111) Ti-terminated}
    \label{fig:111-Titerm-bond-site-stc-rlx}
  \end{subfigure}
  \begin{subfigure}[b]{0.45\linewidth}
    \includegraphics[width=\linewidth]{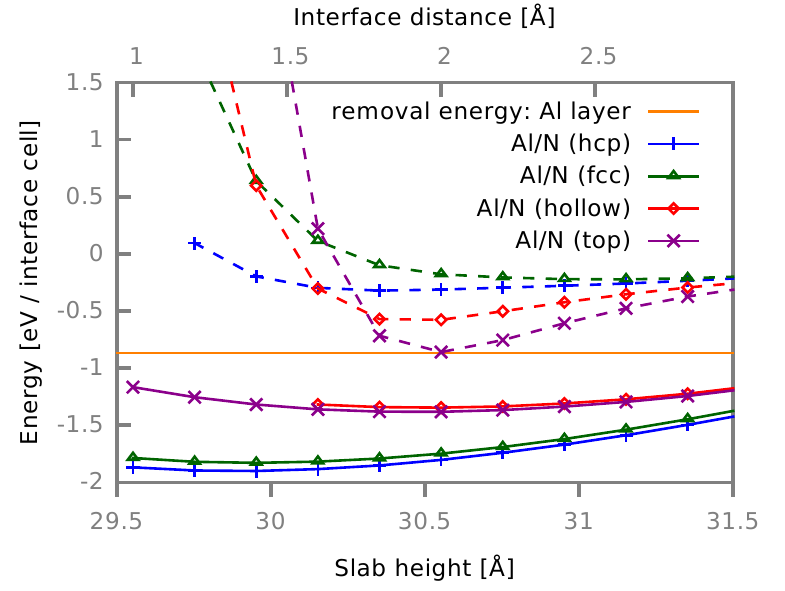}
    \caption{(111) N-terminated}
    \label{fig:111-Nterm-bond-site-stc-rlx}
  \end{subfigure}
  \caption{Calculated PBE interaction energies of the Al/TiN interface for the (111) Ti- and N-terminated  surfaces. Solid and dashed lines indicate results of relaxed and static calculations, respectively. The slab heights on the lower x axis are valid for static and relaxed calculations, whereas the interface distances on the upper x axis refer only to the static calculations. Various lateral alignments of the two slabs are considered (see Fig.~\ref{fig:bond-sites}). The horizontal solid orange lines give the energy costs to remove at least one layer from an Al slab.}
  \label{fig:bond-sites-stc-rlx}
\end{figure*}

Generally, relaxation effects have to be accounted for to obtain the correct equilibrium values of the adhesion energy and the interface distance as well as to predict the occurrence of material transfer. A comparison between the  relaxed and static results is given in Fig.~\ref{fig:bond-sites-stc-rlx} for the (111) surfaces. For rather closed TiN surfaces, such as the (001) and Ti-terminated (111) orientations [see Fig.~\ref{fig:bond-sites-stc-rlx}(a)], relaxations typically cause only small changes in the equilibrium quantities of the interface. Hence, computationally ``cheap'' static calculations give good estimates, unless pronounced changes in the structure of the Al slab occur. This is, for example, the case for the Al/Ti~(hollow) alignment of the (111)~Al/TiN (Ti-terminated) interface, since the interfacial Al atom relaxes towards the energetically more favorable fcc contact site. In the case of the more open (011) surface, relaxations show more pronounced effects for all alignments and should be taken into account. Nevertheless, the energy hierarchy and the prediction of the occurrence of material transfer are not affected for all alignments with the exception of the Al/TiN~(hollow) case. Again, the Al/TiN~(hollow) interface behaves differently because the relaxed structure of the Al slab is modified by the approaching TiN slab. In more detail the interfacial Al layer is moved to the Al/N~(bridge) site, which is the most favorable alignment. This movement of about \unit[0.8]{\AA} occurs mainly in the lateral plane. The free subinterface layers are shifted to gradually compensate the change in the stacking between the fixed layers at the bottom of the slab and the interfacial layer. These shifts range approximately between \unit[0.2]{} and \unit[0.6]{\AA}. For the cases discussed so far, except for the hollow alignments,  relaxations showed rather small effects on the equilibrium quantities. In contrast, all alignments of  the (111)~Al/TiN (N-terminated) interface are crucially affected by relaxations [see Fig.~\ref{fig:bond-sites-stc-rlx}(b)]. The adhesion energies are strongly increased, and the energetical hierarchy of the alignments is altered. Furthermore, while static calculations suggest the absence of material transfer, relaxations predict its occurrence for all tested alignments.

\begin{figure*}[hbt]
  \centering
  \begin{subfigure}[b]{0.45\linewidth}
    \includegraphics[width=\linewidth]{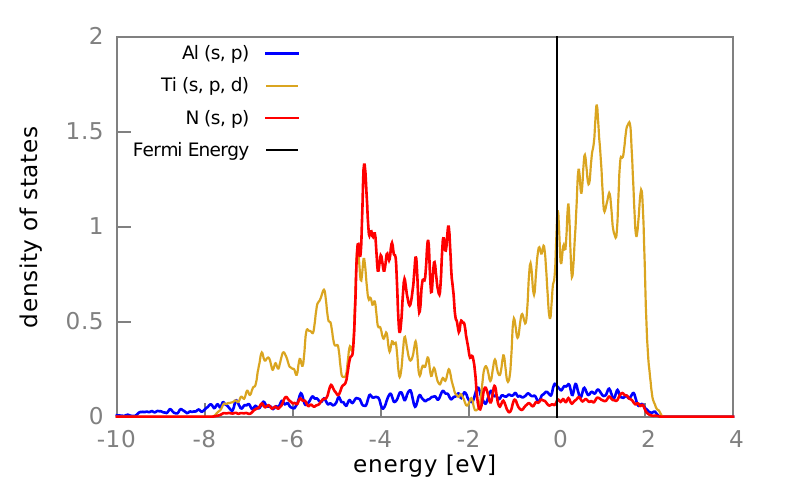}
    \caption{Al \& TiN~(isolated): surface layers}
    \label{fig:DOS-al-tin-iso-l1}
  \end{subfigure}
  \begin{subfigure}[b]{0.45\linewidth}
    \includegraphics[width=\linewidth]{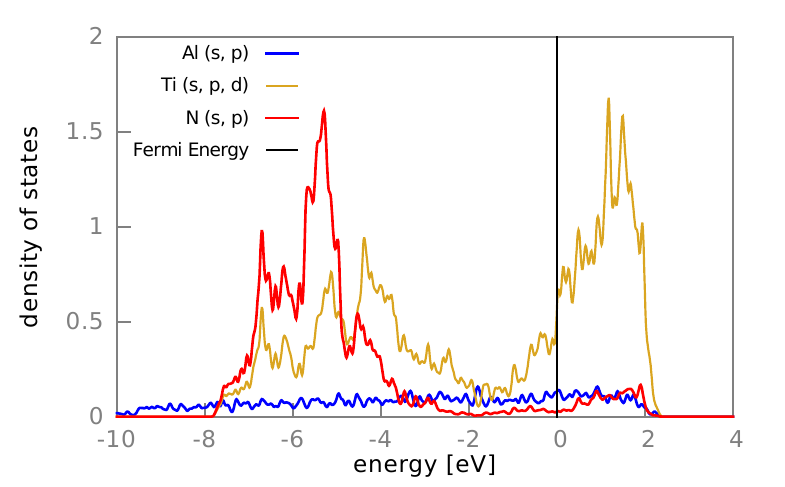}
    \caption{Al \& TiN~(isolated): sub-surface layers}
    \label{fig:DOS-al-tin-iso-l2}
  \end{subfigure}
  \begin{subfigure}[b]{0.45\linewidth}
    \includegraphics[width=\linewidth]{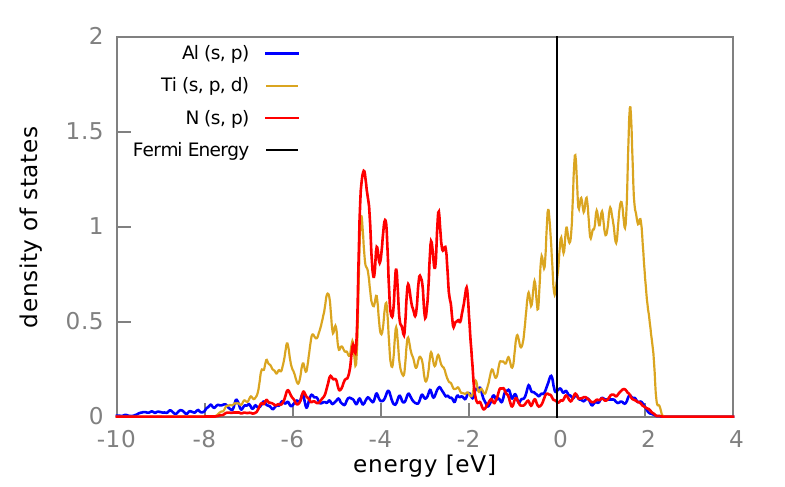}
    \caption{Al/Ti~(top): interface layers}
    \label{fig:DOS-al-t-ti-2.8-l1}
  \end{subfigure}
  \begin{subfigure}[b]{0.45\linewidth}
    \includegraphics[width=\linewidth]{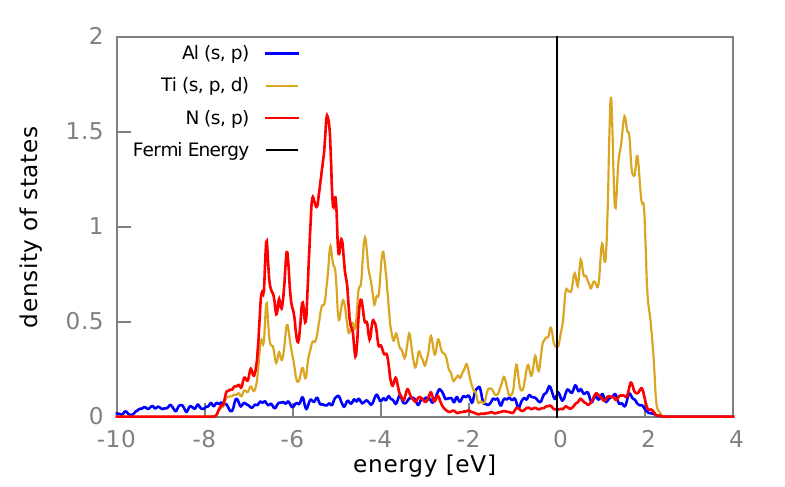}
    \caption{Al/Ti~(top): sub-interface layers}
    \label{fig:DOS-al-t-ti-2.8-l2}
  \end{subfigure}
  \begin{subfigure}[b]{0.45\linewidth}
    \includegraphics[width=\linewidth]{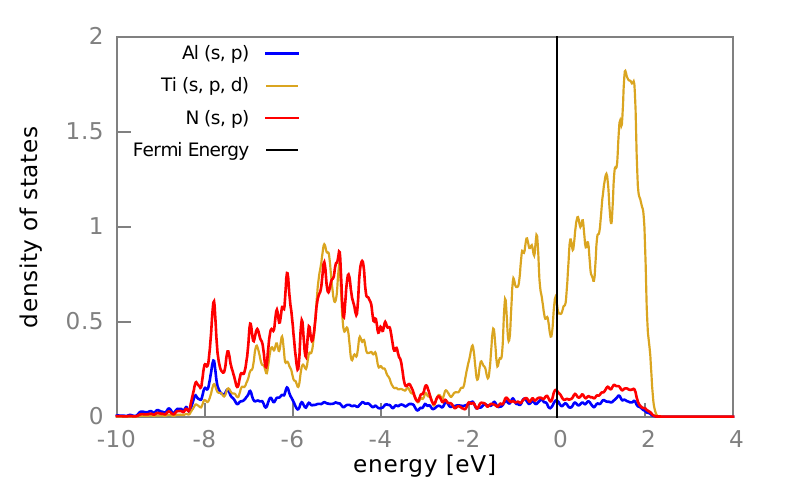}
    \caption{Al/N~(bridge): interface layers}
    \label{fig:DOS-al-b-n-1.8-l1}
  \end{subfigure}
  \begin{subfigure}[b]{0.45\linewidth}
    \includegraphics[width=\linewidth]{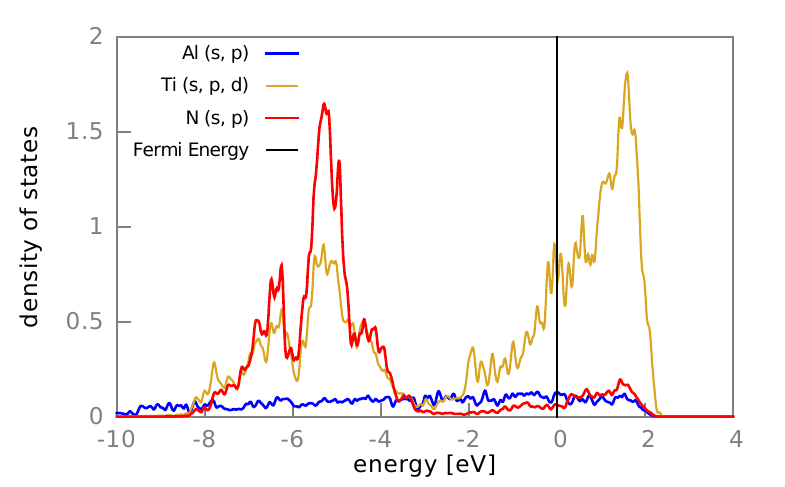}
    \caption{Al/N~(bridge): sub-interface layers}
    \label{fig:DOS-al-b-n-1.8-l2}
  \end{subfigure}
  \caption{Layer-projected DOSs from PBE calculations of the isolated Al and TiN slabs as well as of the (011) Al/TiN interface for the Al/Ti~(top) and Al/N~(bridge) alignments. The Fermi energy is shifted to \unit[0]{eV}.}
  \label{fig:DOS-110}
\end{figure*}

For a better understanding of the energetically preferred configurations at the interface, layer-projected densities of state (DOSs) and differences in charge densities are examined. Layer-projected DOSs are displayed in Fig.~\ref{fig:DOS-110} for the two alignments Al/N~(bridge) and Al/Ti~(top) as well as the isolated slabs of the (011) surface orientation. This surface orientation has been chosen because it exhibits a large spread in adhesion energies for different alignments. Additionally, the occurrence of material transfer should depend on the alignment. In Fig.~\ref{fig:DOS-110} ``interface (surface) layers'' indicate the first layers of Al, Ti, and N immediately at the interface (surface), whereas ``subinterface (subsurface) layers'' mean the next layers of Al, Ti, and N moving deeper into both materials. Further layers are not presented because they exhibit only minor differences with respect to the subinterface layers. The DOSs of the shown alignments display distinct features. For the Al/Ti~(top) case, where Ti is the next interfacial neighbor of the top Al atom, the Al DOS is almost not affected by the interface. Only a small accumulation of sp states just below the Fermi energy and a depletion of s states at the edges of the DOS are found for the interface layers with respect to the other layers. The N sp states are shifted closer to the Fermi energy for the interfacial layer, and in particular, the Ti d states exhibit more occupied states at the Fermi energy. These changes indicate a weakly covalent bonding between the Al sp states and the Ti d states. Furthermore, the DOS is very similar to the case of the isolated Al and TiN slabs. This also shows the weak interaction for the Al/Ti~(top) interface. On the other hand, for the Al/N~(bridge) configuration, where the uppermost Al atoms are closer to N across the interface, the Al DOS is changed in a more pronounced way. The sp states in the interface layers are partially shifted to lower energies, resulting in a pronounced peak at about \unit[-8]{eV} and a few minor ones around \unit[-7]{eV}. The N p states around \unit[-5]{eV} are broadened in the interfacial layer, resulting in common peaks with Al states roughly between \unit[-6]{eV} and \unit[-8]{eV}. These effects at the interface indicate a hybridization of Al and N sp states and explain the stronger adhesion due to covalent interaction. The interfacial Ti states are only slightly affected, exhibiting a few more occupied states at the Fermi level.

\begin{figure}[hbt]
  \centering
  \begin{subfigure}[b]{0.2\linewidth}
    \includegraphics[width=\linewidth]{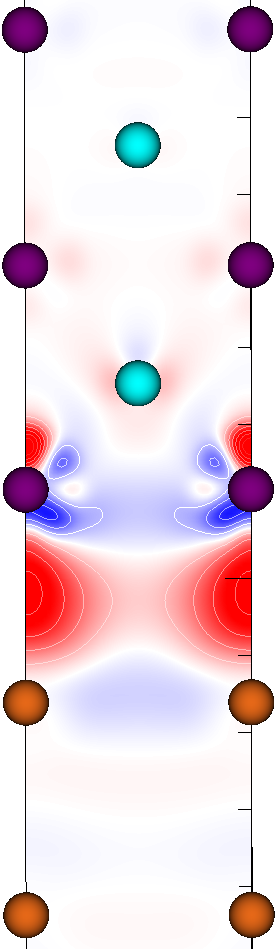}
    \caption{\\ Al/Ti (top): Ti-plane}
    \label{fig:chg-al-t-ti-x100}
  \end{subfigure}
  \hspace{0.02\linewidth}
  \begin{subfigure}[b]{0.2\linewidth}
    \includegraphics[width=\linewidth]{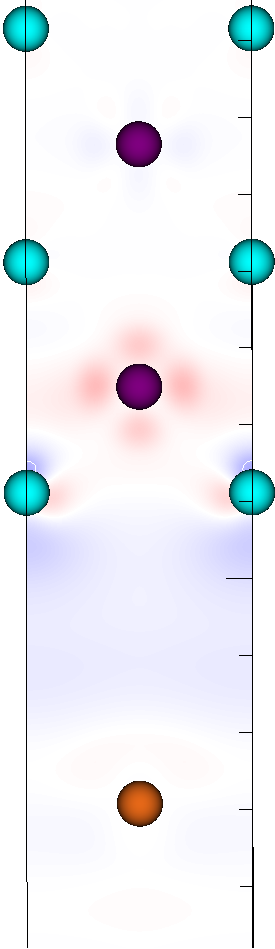}
    \caption{\\ Al/Ti (top): N-plane}
    \label{fig:chg-al-t-ti-x50}
  \end{subfigure}
  \hspace{0.05\linewidth}
  \begin{subfigure}[b]{0.2\linewidth}
    \includegraphics[width=\linewidth]{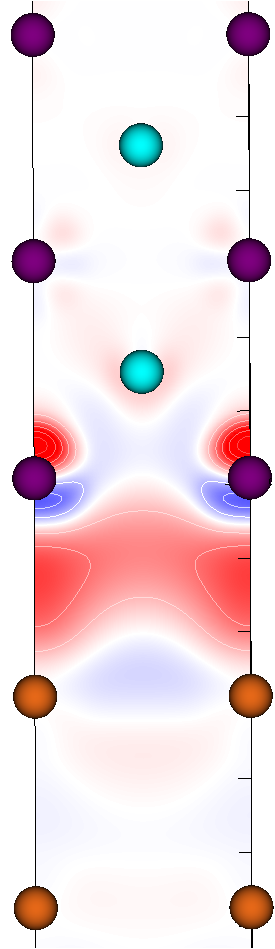}
    \caption{\\ Al/N (bridge): Ti-plane}
    \label{fig:chg-al-b-n-x100}
  \end{subfigure}
  \hspace{0.02\linewidth}
  \begin{subfigure}[b]{0.2\linewidth}
    \includegraphics[width=\linewidth]{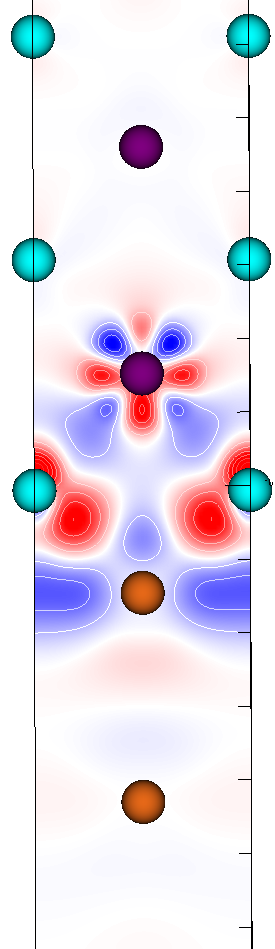}
    \caption{\\ Al/N (bridge): N-plane}
    \label{fig:chg-al-b-n-x50}
  \end{subfigure}
  \caption{Charge-density differences \(\rho_{diff}\) [see Eq.~\eqref{equ:charge-diff}] of the (011) Al/TiN interface. \(\rho_{diff}\) was obtained from PBE calculations for the relaxed equilibrium configurations of (a) and (b) the Al/Ti~(top) alignment and  (c) and (d) the Al/N~(bridge) alignment. The charge-density difference of each alignment is plotted for the Ti plane and the N plane (recall Fig.~\ref{fig:bond-sites}) for values from -0.2 (solid blue, deficit) to 0.2 (solid red, accumulation) electrons/\unit[]{\AA\(^3\)}. Color code: Al, orange; Ti, violet; N, cyan.}
  \label{fig:chg-diff-011}
\end{figure}

In addition to the DOS, charge densities at the interfaces are investigated and presented for the same alignments of the (011) surface. For a better visualization the differences of charge densities \(\rho_{diff}\) between the Al/TiN interface and the isolated, independent Al and TiN slabs are presented in Fig.~\ref{fig:chg-diff-011}. The charge-density difference \(\rho_{diff}\) is defined as
\begin{equation}
  \centering
  \rho_{diff}=\rho_{Al/TiN} - (\rho_{Al} + \rho_{TiN}),
  \label{equ:charge-diff}
\end{equation}
where \(\rho_{Al/TiN}\) is the charge density of the interface, while \(\rho_{Al}\) and \(\rho_{TiN}\) represent the charge densities of the isolated slabs. Both displayed alignments result in a rather continuous charge accumulation between Al and Ti at the interface, suggesting a bonding [see Figs.~\ref{fig:chg-diff-011}(a) and~\ref{fig:chg-diff-011}(c)]. For the Al/N~(bridge) configuration an additional charge buildup occurs between the interfacial Al and N atoms, which indicates covalent contributions to the bonding due to the more localized and directional character of the accumulation [see Fig.~\ref{fig:chg-diff-011}(d)]. These findings support the DOS arguments made in the previous paragraph.

\subsection{Approach and Separation of Al and TiN Slabs}\label{subsec:interface-loop}

\begin{figure*}[hbt]
  \centering
  \begin{subfigure}[b]{0.45\linewidth}
    \includegraphics[width=\linewidth]{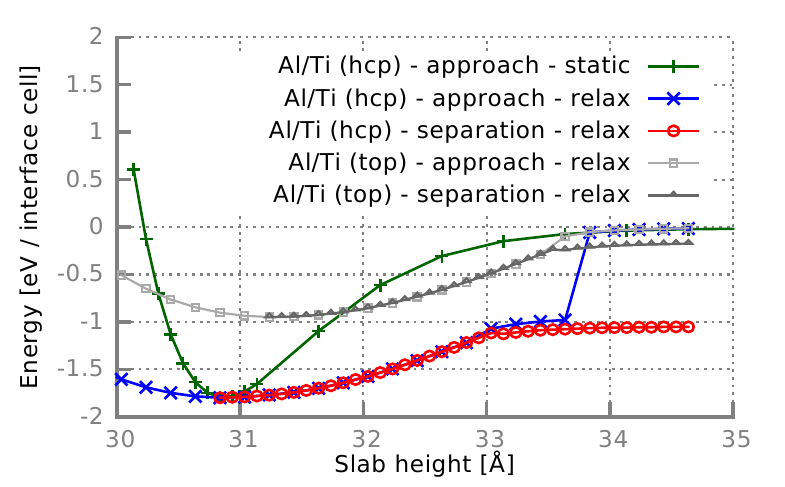}
    \caption{(111) Ti-terminated}
    \label{fig:111-pec}
  \end{subfigure}
  \begin{subfigure}[b]{0.45\linewidth}
    \includegraphics[width=\linewidth]{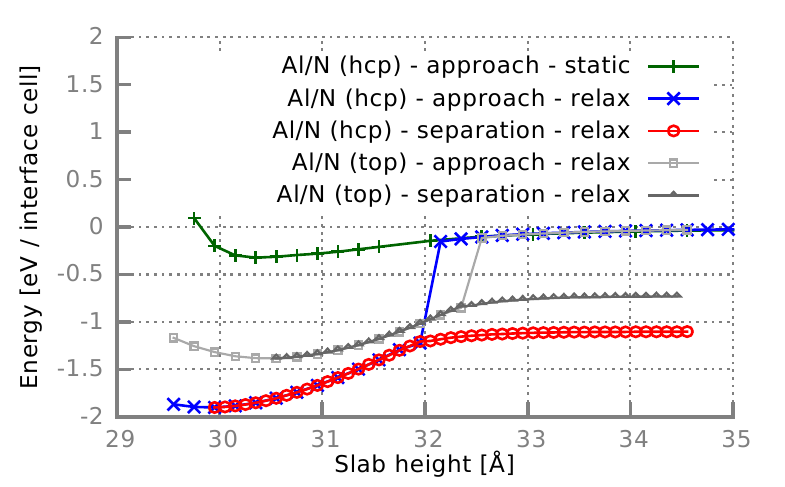}
    \caption{(111) N-terminated}
    \label{fig:111-nterm-pec}
  \end{subfigure}
  \begin{subfigure}[b]{0.45\linewidth}
    \includegraphics[width=\linewidth]{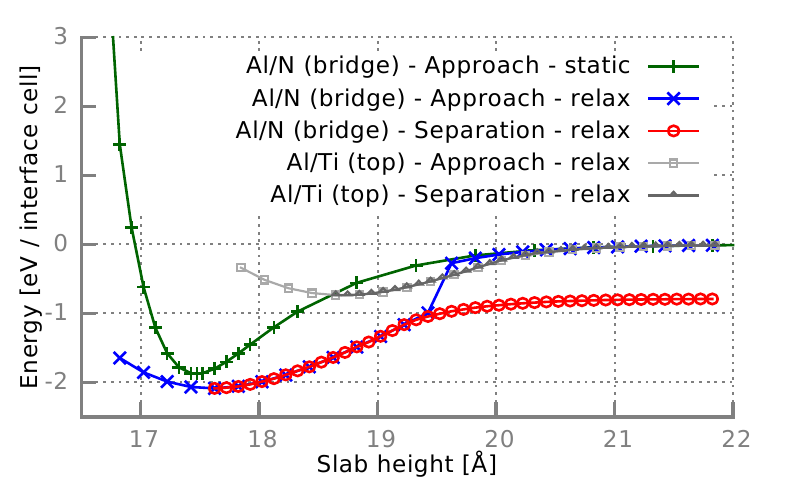}
    \caption{(011)}
    \label{fig:110-pec}
  \end{subfigure}
  \begin{subfigure}[b]{0.45\linewidth}
    \includegraphics[width=\linewidth]{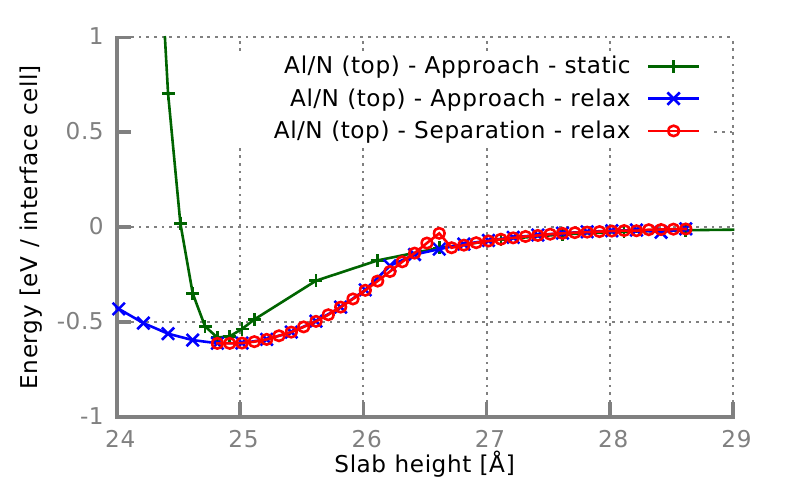}
    \caption{(001)}
    \label{fig:100-pec}
  \end{subfigure}
  \caption{Calculated PBE interaction energies [see Eq.~\ref{equ:interaction-energy}] for the approach and subsequent separation of Al and TiN slabs for (111) Ti-terminated, (111) N-terminated, (011), and (001) surface orientations. The alignments follow the definitions in Fig.~\ref{fig:bond-sites}.}
  \label{fig:pecs}
\end{figure*}

The energetical argument on material transfer presented above can be tested by ``slowly'', i.e., using small discrete steps, approaching and subsequently separating the slabs. The energetical results of such loops are depicted in Figs.~\ref{fig:pecs}(a)--(d) for different configurations; the respective energies are presented in Table~\ref{tab:loop-data}. The green curves with their data points indicated by pluses in Figs.~\ref{fig:pecs}(a)--(d) give static potential-energy curves, where all atoms were kept rigid for each selected interface distance. For large interface distances this curve shows the limiting case of separated, independent slabs. For ever-shorter distances the effect of relaxation becomes important, and the actual energies deviate from the green curves. The blue curves with their data points marked by crosses show the interaction energies of the approaching slabs including atomic relaxations after each discrete step. For some of these cases we find rather large jumps which are either due to material transfer between the slabs, namely, from Al to TiN, or due to the Al slab expanding into the space between the slabs. Finally, the red curves with their data points displayed by circles indicate the interaction energies of the subsequent separation of the slabs, again including atomic relaxations after each step. These curves are also not completely smooth but display some kinks or smaller jumps mainly due to the breaking apart of the Al/TiN slab into two separated ones. When material transfer takes place, these curves, of course, do not approach the green ones, even at large slab separations.

\begin{table*}[hbt]
  \caption{\label{tab:loop-data} Equilibrium interface distances, adhesion energies, energy costs to remove layers from the Al slab, and number of transferred Al layers for various interface configurations. For the (111) orientation Al/Ti and Al/N denote the Ti- and N-terminated surfaces, respectively.}
  \begin{ruledtabular}
    \begin{tabular}{lcccc}
      & Equilibrium  & Adhesion energy & Removal energies & Material transfer \\
      & interface distance [\unit{\AA}] &  [\unit{eV}/interface cell] & [\unit{eV}/interface cell] & [Al layers] \\
      (001) Al/N~(top) & 2.06 & -0.61 & 1.16 & 0 \\
      (011) Al/N~(bridge) & 1.39 & -2.09 & 1.35 & 2 \\
      (011) Al/Ti~(top) & 2.77 & -0.73 & 1.35 & 0 \\
      (111) Al/Ti~(hcp) & 2.22 & -1.78 & 0.80 & 2 \\
      (111) Al/Ti~(top) & 2.67 & -0.94 & 0.80 & 1 \\
      (111) Al/N~(hcp) & 1.04 & -1.90 & 0.80 &  2 \\
      (111) Al/N~(top) & 1.87 & -1.38 & 0.80 & 1 
    \end{tabular}
  \end{ruledtabular}
\end{table*}

\begin{figure}[hbt]
  \centering
  \begin{subfigure}[b]{0.167\linewidth}
    \includegraphics[width=\linewidth]{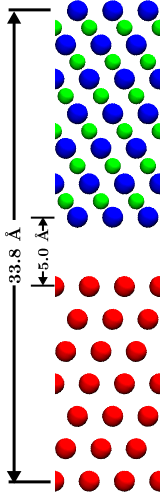}
    \caption{\\ \unit[33.8]{\AA}}
    \label{fig:111-in1}
  \end{subfigure}
  \hspace{0.02\linewidth}
  \begin{subfigure}[b]{0.11\linewidth}
    \includegraphics[width=\linewidth]{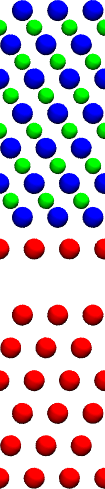}
    \caption{\unit[33.6]{\AA}}
    \label{fig:111-in2}
  \end{subfigure}
  \hspace{0.02\linewidth}
  \begin{subfigure}[b]{0.11\linewidth}
    \includegraphics[width=\linewidth]{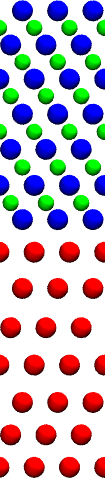}
    \caption{\unit[32.8]{\AA}}
    \label{fig:111-in3}
  \end{subfigure}
  \hspace{0.02\linewidth}
  \begin{subfigure}[b]{0.11\linewidth}
    \includegraphics[width=\linewidth]{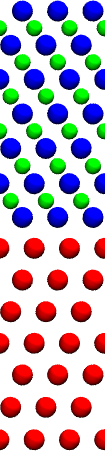}
    \caption{\unit[30.8]{\AA}}
    \label{fig:111-in4}
  \end{subfigure}

  \vspace{0.5cm}
  \begin{subfigure}[b]{0.167\linewidth}
    \includegraphics[width=\linewidth]{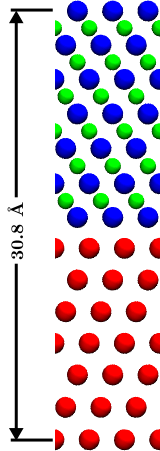}
    \caption{\\ \unit[30.8]{\AA}}
    \label{fig:111-out1}
  \end{subfigure}
  \hspace{0.02\linewidth}
  \begin{subfigure}[b]{0.11\linewidth}
    \includegraphics[width=\linewidth]{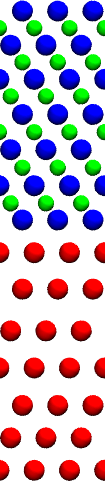}
    \caption{\unit[33.0]{\AA}}
    \label{fig:111-out2}
  \end{subfigure}
  \hspace{0.02\linewidth}
  \begin{subfigure}[b]{0.11\linewidth}
    \includegraphics[width=\linewidth]{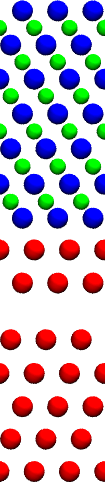}
    \caption{\unit[33.1]{\AA}}
    \label{fig:111-out3}
  \end{subfigure}
  \hspace{0.02\linewidth}
  \begin{subfigure}[b]{0.11\linewidth}
    \includegraphics[width=\linewidth]{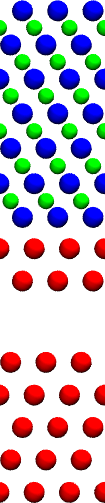}
    \caption{\unit[34.6]{\AA}}
    \label{fig:111-out4}
  \end{subfigure}
  \caption{(a)--(d) Approach and (e)--(h) separation of (111) Al and (111) TiN slabs. Al, Ti and N are colored in red, blue, and green, respectively. (d) and (e) show the structure at the relaxed equilibrium distance. The steps are defined by the slab height.}
  \label{fig:111-app-sep}
\end{figure}

For the Ti-terminated (111) surface orientation potential-energy curves are presented in Fig.~\ref{fig:pecs}(a) for the two extremal alignments, Al/Ti~(hcp) and Al/Ti~(top), which show the highest and lowest adhesion energies. As expected from the energetics, material transfer occurs during separation, and both systems end up in an energetically more favorable configuration compared to the initial setup. In particular, one and two Al layer(s) for Al/Ti~(top) and Al/Ti~(hcp), respectively, are transferred. This discrepancy in the number of transferred layers cannot be explained from the energetics but could stem from the different equilibrium interface distances. Compared to that of the hcp alignment, this distance is significantly increased by almost 20\% for the top configuration, hindering the interaction between TiN and the subinterface Al layer. For the Al/Ti~(hcp) configuration snapshots of the structures during approach and separation are presented in Fig.~\ref{fig:111-app-sep}. During the approach at a slab height of about \unit[33.6]{\AA} a large drop in interaction energy occurs for the Al/Ti~(hcp) alignment [see Fig.~\ref{fig:pecs}(a)] due to material transfer of the topmost Al layer to the TiN slab [see Fig.~\ref{fig:111-app-sep}(b)]. This is not the ground state since a transfer of two layers would yield an even lower total energy. At this distance, the transfer of the second Al layer is hindered by an energy barrier of about \(E_{b2} \approx\)~\unit[324]{meV}, which is significantly larger than for the first layer alone, \(E_{b1} \approx\)~\unit[122]{meV}. Upon further approaching, at a slab height of about \unit[32.8]{\AA}, a slight kink occurs [see Fig.~\ref{fig:pecs}(a)] because the Al slab is expanded into the space between the slabs [see Fig.~\ref{fig:111-app-sep}(c)]. For a further approach, the interaction energy follows an essentially parabolic curve until the minimum energy is reached [see Figs.~\ref{fig:pecs}(a) and~\ref{fig:111-app-sep}(d)]. The subsequent separation is started from the equilibrium structure at the interaction energy minimum. At first the red interaction energy curve lies on top of the blue one [see Fig.~\ref{fig:pecs}(a)], meaning that the Al slab becomes extended again [see Fig.~\ref{fig:111-app-sep}(f)]. At a slab height of about \unit[33.1]{\AA} the two curves for approach and separation start to deviate [see Fig.~\ref{fig:pecs}(a)] when the Al/TiN compound separates [see Fig.~\ref{fig:111-app-sep}(g)]. Two Al layers stick to the TiN slab and form a stable configuration. This behavior during the complete loop is typical for all cases exhibiting material transfer. While the Al slab is strongly affected by the approach of the TiN slab, almost no changes in the TiN structure are observed. The more pronounced impact on the Al slab is not surprising when considering that TiN forms a much more rigid lattice than Al. This claim is  not entirely valid for the N-terminated (111) TiN slab, which will be discussed in the following paragraph. Using the finally stable state (TiN plus two Al layers) as a starting configuration for a new loop of approach and separation versus an Al slab yields a reversible cycle. This should be kept in mind when one is interpreting, for example, AFM experiments. Upon the first contact between the tip and a particular spot on a surface material transfer might occur, which in turn changes the contact properties and forces between the tip and the surface. However, further encounters on the same spot should then be within the reversible cycle and lead to the same response. 

The N-terminated (111) orientation is, in some respects, very similar to the Ti-terminated one. As predicted, both configurations yield material transfer for all tested alignments (see Fig.~\ref{fig:pecs}). However, as explained above, static calculations completely fail to describe the equilibrium quantities of the N-terminated case, whereas they result in good estimates for the other orientations. This discrepancy is due to the behavior of the interfacial N layer for N-terminated (111) TiN. In the absence of the counter Al slab, the surface N layer is closely bound to the next Ti layer at a distance of \unit[0.84]{\AA}, while in contact with an Al slab the distance grows to \unit[1.47]{\AA} at the equilibrium configuration. This behavior is crucial for the energetics and can be captured only when relaxations are included. The interfacial N layer is actually closer to the next Al layer with a distance of \unit[1.11]{\AA} than to the next Ti one. Due to this result the possibility of a diffusion of the interfacial N layer into the Al slab was investigated. For all alignments with the exception of Al/N (top), no energetically favorable configurations were found. However, for the Al/N (top) alignment the exchange of the interfacial Al and N layers and a   subsequent relaxation of the system yield a favorable state by about \unit[683]{meV}, which is also about \unit[235]{meV} lower than the previously found minimum for the Al/N (hcp) alignment. In this favorable configuration an Al-N-Al trilayer is formed, showing the wurtzite structure, which is typically observed in aluminum nitride crystals. From the thermodynamical point of view diffusion seems to be possible. Of course, for the full picture reaction paths and energy barriers have to be considered. 

The (011) surface orientation also presents an interesting case because due to the energetic results [see Fig.~\ref{fig:bond-sites-pec}(c)], material transfer is expected for only the alignments Al/N~(bridge), Al/N~(top), and Al/TiN~(hollow). As an example, one can see from the loops given in Fig.~\ref{fig:pecs}(c) that Al/N~(bridge) shows a favorable configuration after separation corresponding to the transfer of two Al layers, whereas the Al/Ti~(top) case is reversible upon approach and separation without any material transfer.

Finally, for the (001) surface orientation material transfer is not expected for any of the alignments. Among all cases Al/N~(top) has the largest adhesion energy; therefore, if a material transfer occurs, it will happen for this case.  However, since the energy cost for the removal of an Al layer exceeds the adhesion energy, no material transfer is observed [Fig.~\ref{fig:pecs}(d)].  The deviation of the curves for approach and separation around \unit[26.5]{\AA} slab height occurs due to the expansion of the Al slab upon separation until the interface breaks apart and relaxes into the initial Al and TiN slabs.

In the literature some publications on tensile test simulations of Al/TiN interfaces can be found, where the separation is achieved by increasing the size of the whole simulation cell in one direction in discrete steps including interim relaxations. Liu et al.~\cite{liu_first-principles_2005} and Zhang et al.~\cite{zhang_first-principles_2007} investigated the Al/TiN (111) and (001) interfaces, respectively. Liu et al. obtained similar results with respect to material transfer for the hcp alignment at the (111) surface for both terminations but did not examine any further alignments of Al and TiN slabs at the interface. Zhang et al. studied the Al/N~(top) configuration of the (001) interface and, in contrast to our work, found a material transfer of the top Al layer. This discrepancy could stem from the different simulation approaches and computational details. However, we repeated these calculations using a setup for the separation of the slabs similar to that of Zhang et al. and did not find any material transfer. Additional simulations for the different setups used by Zhang et al. and in the present investigation testing the influence of varying step sizes also did not lead to a material transfer.

\subsection{Comparison of Surface Energies}\label{subsec:surf-energy}

The behavior of the different surface orientations can also be discussed from the surface energy's point of view. The surface energies for Al and TiN slabs are presented in Table~\ref{tab:surf-energy}.  It has to be noted that for (111) TiN the surface energy depends on the termination and the chemical potential of nitrogen. Here the lowest possible value for the surface energy is used, which is achieved by the N-terminated surface at \(\Delta\mu_N = 0\) (see Fig.~\ref{fig:TiN-term}). The value of the surface energy for the Ti-terminated surface at \(\Delta\mu_N = 0\) is about three times larger, but its minimum is comparable to that of the N-terminated case. For the (001) and (011) orientations the surface energy is independent of the chemical potential~\cite{wang_surface_2010}. As shown in Table~\ref{tab:surf-energy}, the Al surfaces always exhibit a smaller surface energy than the TiN ones. The differences between Al and TiN are pronounced for the (011) and (111) orientations. For (001), however, the surface energies are rather comparable. Material transfer can be seen as a measure of surface-energy minimization by creating a new energetically ``cheap'' surface and covering an ``expensive'' one with it. This argument provides a hint about which surface orientations may favor material transfer. However, for the full picture also other contributions such as the interaction energy, which is influenced additionally by the alignment of the slabs, also have to be considered. For example, the surface-energy argument would suggest the possibility of material transfer for (001) and cannot explain why only some (011) configurations exhibit this feature. 
\begin{table}[hbt]
  \caption{\label{tab:surf-energy} Surface energies (in \unit{eV/\AA\(^2\)}) of the Al and TiN slabs for the (001), (011), and (111) surface orientations. In the case of the (111) TiN surface the N-terminated one at \(\Delta\mu_N = 0\) is given here because it exhibits the lowest surface energy of all (111) TiN surfaces (see Fig.~\ref{fig:TiN-term}).}
  \begin{ruledtabular}
    \begin{tabular}{lccc}
      & (001) & (011) & (111) \\
      Al & 0.058 & 0.064 & 0.052 \\
      TiN & 0.087 & 0.174 & 0.094 
    \end{tabular}
  \end{ruledtabular}
\end{table}

\subsection{Assessment of Computed Results}\label{subsec:add-tests}

To validate the information presented above various additional tests were performed. The results will be presented for the Ti-terminated (111) Al/Ti~(hcp) configuration. First of all, finite-size effects are a major concern. Thus, the size of the simulation cell was increased laterally up to a 3\(\times\)3 surface cell and vertically up to a 19-layer Al slab. Also, intermediate Al slab thicknesses were examined. The TiN slab was not extended vertically because it is almost not affected by the approach of the Al slab. In the case of laterally magnified simulation cells the number of \textbf{k} points was decreased accordingly, e.g., for a 3\(\times\)3 surface cell, a 5\(\times\)5\(\times\)1 mesh was used. For all tested systems the equilibrium interface distances, adhesion energies, and energy costs to remove Al layers were found within about 2\% of the values given above. The energies are referenced to 1\(\times\)1 surface cells. Particularly, the results on material transfer were not affected, meaning that the number of transferred Al layers was not altered.  For the 3\(\times\)3 surface cell the effect of fluctuations at the surface was tested by moving one atom out of the surface plane at several interface distances before material transfer occurs. These tests resulted in the transfer of entire layers too since the shifted atom either relaxed back into its originating slab or was transferred together with the rest of the layer.

\begin{figure}[hbt]
  \centering
  \includegraphics[width=0.9\linewidth]{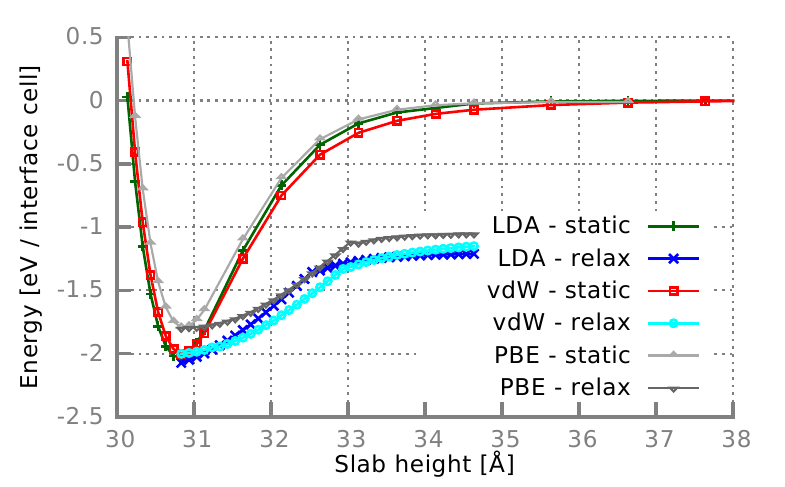}
  \caption{Comparison of calculated interaction energies for the Ti-terminated (111) Al/Ti~(hcp) configuration and various exchange correlation functionals, namely, PBE, LDA, and opbB86b. The relaxed energies represent the separation of the slabs starting from the equilibrium configuration. The differences between static and relaxed curves at large heights occur due to material transfer.}
  \label{fig:111-vgl-xc}
\end{figure}

Furthermore, the effect of the chosen lattice parameters was investigated. The simulations were repeated using the lattice constants of pure Al and TiN for the lateral lattice parameters of the simulation cell. The equilibrium interface distance again changed by only about 2\%. Although the adhesion energies were altered by about 4\%, the removal energies were affected in a similar way, resulting in the same material transfer. Moreover, the influence of other approximations for the exchange correlation functional was tested as already discussed for the removal energies above. The results are presented in Fig.~\ref{fig:111-vgl-xc}. The adhesion energies were enhanced similar to the removal energies, again producing the same results for material transfer. Using the vdW functional, the interaction between the slabs started at larger interface distances. This behavior is expected because of the nonlocal correction added in the vdW functional.

\section{Conclusion and Outlook}\label{sec:conclusion}

Al/TiN interfaces were examined in detail by investigating the contact between Al and TiN slabs showing low-index surface orientations within the framework of density functional theory. Moreover, these contacts were established for various lateral alignments of the slabs at the interface. It was shown that interfacial properties such as the adhesion energy and the equilibrium structure sensitively depend on the given configuration. This behavior can be qualitatively explained by comparing the densities of state and the charge densities of different configurations because distinct bond situations are revealed at the interface. Furthermore, the approach and subsequent separation of Al and TiN slabs were simulated to study the effect on the slabs, especially the possibility of material transfer. The transfer of material from an Al toward a TiN slab was observed for interfacial configurations, which exhibited a larger adhesion energy than the energy cost to remove layers from the Al slab. This is in agreement with the observation that metal-ceramic interfaces break at the interface or in bulk areas according to their interfacial adhesion~\cite{howe_bonding_1993-1, ernst_metal-oxide_1995}. The removal energy for Al layers was found to depend on tensile or compressive stress. In all systems showing material transfer one or two layers of Al stick to the TiN slab after the separation and form an energetically favorable compound with respect to the initial configuration. The differences in surface energies between the slabs are not sufficient to explain the occurrence of material transfer because the given alignment at the interface has to be considered as well. All results were tested for various computational setups such as different sizes of the investigated system or several approximations for the exchange correlation functional. While properties such as the removal and adhesion energies depend on these settings to some degree, the results for material transfer are not affected.

The method used in this work can be, in principle, applied to any pair of materials. However, complex materials or pairs with an unfavorable bulk lattice mismatch may need very large simulation cells to be considered, which means high computational demands, in order to preserve the translational symmetry and to keep the distortions at an acceptable level. Furthermore, larger cells also allow the inclusion of additional features. For example, the distortions due to the lattice mismatch can be minimized, dislocations as well as quasi-incommensurate contacts can be modeled, and even roughness could be included to some degree, e.g., by using stepped surfaces or a regular grid of small asperities.

\section*{Acknowledgments}\label{sec:acknow}
The authors thank G. Vorlaufer for fruitful discussions. G.F., M.W., P.O.B., P.M. and J.R. acknowledge the support by the Austrian Science Fund (FWF): F4109 SFB ViCoM. Part of this work was funded by the the Austrian COMET-Program (project K2 XTribology, No. 824187 and No. 849109) via the Austrian Research Promotion Agency (FFG) and  the Province of Nieder\"osterreich, Vorarlberg and Wien. This work has been carried out within the ``Excellence Center of Tribology'' (AC2T research GmbH) and at the Vienna University of Technology. Part of this work was supported by the European Cooperation in Science and
Technology (COST; Action MP1303). The authors also appreciate the ample support of computer resources by the Vienna Scientific Cluster (VSC). Figs.~\ref{fig:Al-TiN-initial} and \ref{fig:bond-sites} were created employing \textsc{VESTA}~\cite{momma_vesta_2011}, Fig.~\ref{fig:chg-diff-011} utilizing \textsc{VisIt}~\cite{childs_visit:_2012} and Fig.~\ref{fig:111-app-sep} using \textsc{VMD}~\cite{humphrey_vmd:_1996}.

\bibliography{refs.bib}

\end{document}